\documentclass[rnote,oldversion]{aa}  
\usepackage{graphicx}
\usepackage{natbib}
\newcommand{\HI}{H{\,\small I}}
\newcommand{\paperI}{Paper{\,\small I}}

\newcommand{\HII}{H{\,\small II}}

\usepackage{txfonts}
\begin{document}
   \authorrunning{Emonts et al.}
   \titlerunning{Deep optical imaging of radio galaxy B2~0648+27}
   \title{From major merger to radio galaxy: low surface-brightness stellar counterpart to the giant H{\small \ }I ring around B2~0648+27}

   \subtitle{}

   \author{B.H.C. Emonts
          \inst{1}
          \and
          R. Morganti\inst{2,3}
          \and
          J.H. van Gorkom
          \inst{1}
          \and
          T.A. Oosterloo
          \inst{2,3}
          \and
          E. Brogt\inst{4}
          \and
          C.N. Tadhunter\inst{5} 
           }

   \offprints{B. Emonts}

   \institute{Department of Astronomy, Columbia University,  Mail Code 5246, 550 West 120th Street, New York, N.Y. 10027, USA\\
\email{emonts@astro.columbia.edu}
         \and
Netherlands Foundation for Research in Astronomy, Postbus 2, 7990 AA Dwingeloo, the Netherlands
         \and
Kapteyn Astronomical Institute, University of Groningen, P.O. Box 800, 9700 AV Groningen, the Netherlands
         \and
Steward Observatory, Department of Astronomy, The University of Arizona, 933 North Cherry Avenue, Tucson, AZ 85721, USA
         \and
Department of Physics and Astronomy, University of Sheffield, Sheffield S3 7RH, UK
             }

   \date{}

 
\abstract{We present the detection of a low surface-brightness stellar counterpart to an enormous (190 kpc) ring of neutral hydrogen (\HI) gas that surrounds the nearby radio galaxy B2~0648+27. This system is currently in an evolutionary stage between major merger and (radio-loud) early-type galaxy. In a previous paper we investigated in detail the timescales between merger, starburst and AGN activity in B2~0648+27, based on its unusual multi-wavelength properties (large-scale \HI\ ring, dominating post-starburst stellar population and infra-red luminosity). In  this {\sl Research Note} we present deep optical B- and V-band imaging that provides further evidence for the merger origin of B2~0648+27. The host galaxy shows a distorted optical morphology and a broad tidal arm is clearly present. A low surface-brightness stellar tail or partial ring curls around more than half the host galaxy at a distance of up to 55 kpc from the centre of the galaxy, following the large-scale, ring-like \HI\ structure that we detected previously around this system. The gas and stars in this ring are most likely tidally expelled material that slowly fell back onto the host galaxy after the merger event. There also appear to be sites of star formation in the \HI\ ring that may have formed within the gaseous tidal debris after the merger. We argue that the observed properties of the gas and stars in B2~0648+27, as well as the apparent time-delay between the merger and the starburst event, may be the logical result of a merger between two gas-rich disk galaxies with a prominent bulge, or of a merger between an elliptical and a gas-rich spiral galaxy. There also appears to be a significant time-delay between the merger/starburst event and the current episode of radio-AGN activity.}


   \maketitle
%

\section{Introduction}
\label{sec:intro}

   \begin{figure*}[htb]
   \centering
   \includegraphics[width=0.8\textwidth]{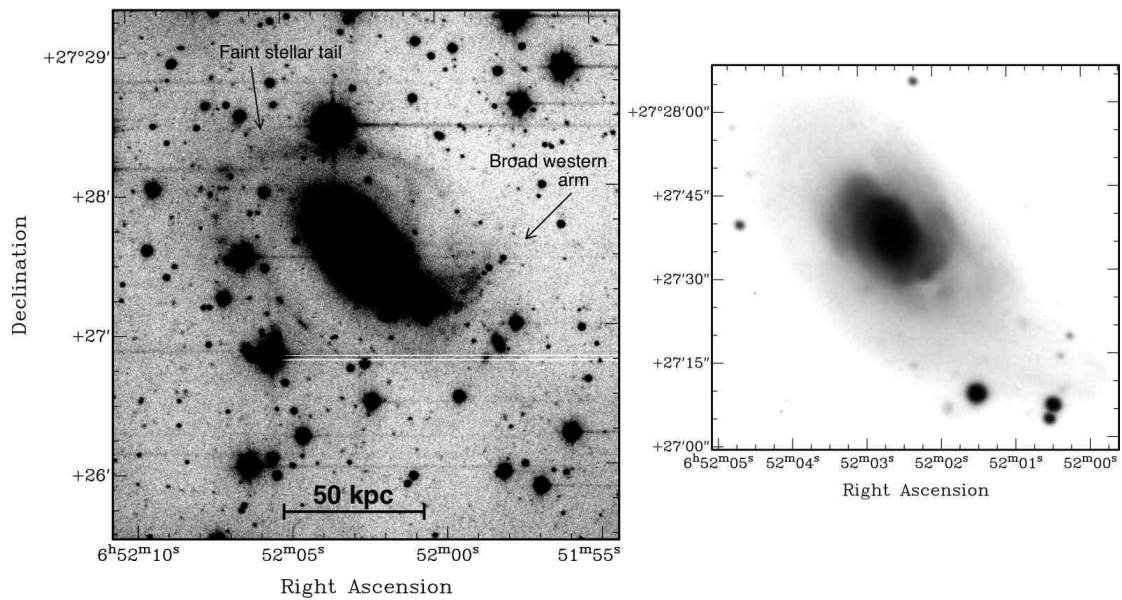}
      \caption{Deep optical B+V band image of radio galaxy B2~0648+27. {\sl Left:} high-contrast plot of B2~0648+27, highlighting the low surface-brightness stellar tail or partial ring. {\sl Right:} low-contrast plot (logarithmic intensity scale) of the inner region of B2~0648+27.}
   \label{fig:optical}
   \end{figure*}

Gas-rich galaxy mergers and interactions are often invoked to be the trigger of starburst and Active Galactic Nuclear (AGN) activity. Gravitational and hydrodynamical perturbations that are exerted on the inter-stellar medium in merging galaxies induce bursts of star formation and may also remove enough angular momentum from the gas for it to be deposited onto a super-massive black hole in the centre of the merging system \citep{bar91,bar96,mih94,mih96,spr05,kap05,dim07}. After a luminous infra-red phase, in which starburst activity is heavily obscured \citep[e.g.][]{san96}, feedback from stellar winds and AGN activity may clear the merger product \citep[e.g.][]{hec90,rup02,dim05,spr05,hop05}. What remains is often a newly formed early-type galaxy. The presence of an active nucleus in such a system may depend critically on the triggering time-scale and life-cycle of the AGN.

In a recent paper \citep[][$\ $hereafter \paperI]{emo06} we investigated in detail the timescales of merger, starburst and AGN activity in the nearby radio galaxy B2~0648+27. Currently, the AGN in B2~0648+27 consists of an emission-line AGN \citep{emo06thesis} plus compact radio source \citep{gir05a}. Although the AGN is not as powerful as often observed in high-redshift quasars and QSOs, the proximity of B2~0648+27 ($z = 0.0412$) allows a detailed study of the host galaxy. In \paperI\ we presented the detailed analysis of a large-scale ring of neutral hydrogen (\HI) gas ($M_{\rm HI} \approx 8.5 \times 10^{9} M_{\odot}$; diameter $\approx$ 190 kpc) that is present around the host galaxy and which was initially discovered by \citet[][]{mor03}. We argued that the \HI\ ring formed from a major merger event $\gtrsim 1.5$ Gyr ago, when gas that was tidally expelled during the merger fell back onto the host galaxy and settled in regular rotation. This merger scenario was supported by the detection of a 0.3-0.4 Gyr young post-starburst stellar population that dominates the starlight throughout the host galaxy and which could have given B2~0648+27 the appearance of an (Ultra-) Luminous Infra-Red Galaxy at the time of the starburst event. The off-set in time between the initial merger event and the starburst activity indicates that the starburst was triggered in an advanced stage of the merger. The spectral age of the radio source in B2~0648+27 has been estimated by \citet{gir05a} to be about 1 Myr, which suggests that there is also a significant time-delay between the starburst event and the triggering of the current episode of radio-AGN activity.

In order to unambiguously verify the merger scenario that we proposed for B2~0648+27 in \paperI\ and to investigate whether stars are present in the large-scale \HI\ ring that surrounds the host galaxy, we obtained deep optical B- and V-band imaging of this system. Previous optical imaging of B2~0648+27 by \citet{hei94} already showed that -- while the host galaxy appears to contain smooth but very elongated elliptical isophotes -- low surface-brightness structures (a faint tail/plume and shells/arcs) appear to be present, although the limited sensitivity of their data only allowed an ambiguous classification of the optical morphology of the host galaxy. A detailed analysis of the radial light-profile of B2~0648+27 from V-band surface photometry by \citet{gon00} revealed that the host galaxy can be fitted by an r$^{1/4}$-law, although isophote twisting is present. This indicates that B2~0648+27 already has a primary optical characteristic of an early-type galaxy, possibly in the process of formation. The new, much deeper optical imaging presented in this {\sl Research Note} clearly shows some remarkable stellar features, which provide additional evidence that a major merger indeed formed this radio galaxy and which allow us to verify our conclusions from \paperI.\\
\vspace{0mm}\\
Throughout this paper we will use the name B2~0648+27 for both the radio source and the host galaxy. We assume $H_{\circ} = 71\ {\rm km~s}^{-1}\ {\rm Mpc}^{-1}$, which puts B2~0648+27 at a distance of 174 Mpc and 1 arcsec = 0.84 kpc.

\section{Observations}
\label{sec:observations}

Deep optical B- and V-band imaging of B2~0648+27 was obtained on 12 (V) and 13 (B) March 2007 at the Hiltner 2.4m telescope of the MDM observatory, located at the south-western ridge of Kitt Peak, Arizona (USA). The total integration times were 60 and 55 minutes for the B- and V-band image respectively. Observations were done under photometric conditions, at low airmass (sec $z < 1.08$) and with a seeing of about 1.3 arcsec on 12 March and about 1.0 arcsec on 13 March. We used the Image Reduction and Analysis Facility (IRAF) to perform a standard data reduction (bias subtraction, flat-fielding, frame alignment and cosmic-ray removal). Probably due to minor shutter issues, a gradient was present in the background of the $9 \times 9$ arcmin Echelle CCD images. Because galaxy B2~0648+27 occupies only a small part of the CCD, we were able to remove this effect to a significant degree by fitting a gradient to the background in the region surrounding B2~0648+27 and subsequently subtracting this background-gradient from the image. However, the residual errors in the background subtraction (visible in Fig. \ref{fig:optical}), in combination with uncertainties in the relative flux calibration, make it impossible to obtain reliable colour information from the B- and V-band images, in particular in the low-surface brightness regions. Using KARMA, we applied a world coordinate system to the images by identifying a few dozen of the foreground stars in an SDSS image of the same region. The newly applied coordinate system agrees with that of the SDSS image to within 1 arcsec. This is good enough for comparing the optical with the \HI\ data, since the latter have a much lower resolution (see \paperI). In order to increase our sensitivity for mapping the very faint optical emission, we co-added the B- and V-band exposures (see Fig. \ref{fig:optical}), although we note that all stellar features were detected in both bands separately.

\section{Results}
\label{sec:results}

\begin{figure*}[t]
   \centering
   \includegraphics[width=0.70\textwidth]{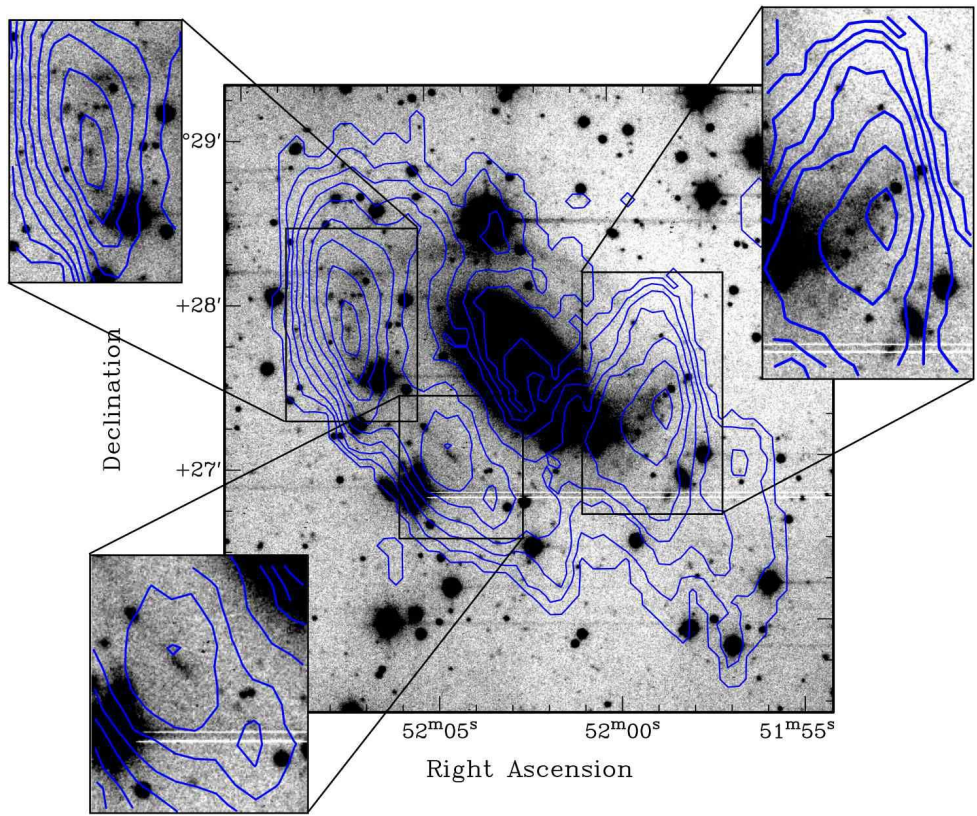}
      \caption{Contours of the total intensity \HI\ map from \paperI\ overlaid onto the deep optical B+V image of B2~0648+27 (grey-scale). Contour levels: 0.22, 0.36, 0.52, 0.71, 0.95, 1.2, 1.5, 1.8, 2.1 $\times 10^{20}$ cm$^{-2}$.}
   \label{fig:overlay}
\end{figure*}

Figure \ref{fig:optical} shows the deep optical B+V band image of radio galaxy B2~0648+27. The high-contrast plot ({\sl left}) clearly shows that the main stellar body of the host galaxy has a total extent of 55 kpc along the major axis. A broad stellar arm appears to extend from the south-western end of the optical galaxy in western direction \citep[this arm was also weakly visible in the R-band image of][]{hei94}. In addition, a low surface-brightness stellar tail or partial ring curls around the northern and eastern half of the galaxy at about 55 kpc from the centre.

Fig. \ref{fig:overlay} shows that the low surface-brightness stellar distribution clearly follows the contours of the \HI\ emission that we detected around B2~0648+27 (see \paperI). A concentration of \HI\ gas is detected on top of the broad stellar arm that stretches west of the optical galaxy. Also the faint stellar tail that curls around the host galaxy follows the \HI\ gas in the ring. At the locations where the \HI\ peaks, enhancements can be seen in the optical structure, possibly indicating sites of star formation (as already mentioned in Sect. \ref{sec:observations}, due to observational limitations no reliable colour information can be obtained from our B- and V-band images in these regions). In a small region between the broad western arm and the north-western tip of the faint stellar tail there is an apparent lack of optical emission. This is also the approximate location where the surface density of the \HI\ ring reaches a minimum. Contrary, no optical emission is detected in the low-surface brightness \HI\ gas that stretches toward the south-west.

The low-contrast plot of Fig. \ref{fig:optical} ({\sl right}) shows in more detail the inner region of the host galaxy. From this plot it is clear that the optical morphology is heavily distorted. Such a distorted optical morphology is typical for the advanced stages of a major merger \citep[e.g.][]{bar96,mih96}. Only one nucleus is visible in our data. This is in agreement with higher resolution {\sl HST} data by \citet{cap00}, which also reveal a single nucleus in an inner region with irregular morphology, enhanced by the presence of dusty features. The {\sl HST} data of \citet{cap00} also reveal a chain of what appear to be star forming regions north of the nucleus, which are aligned parallel to the dusty filaments. The resolution of our optical data is too low to distinguish these individual dust features and star forming regions.



\section{Discussion}
\label{sec:discussion}

Our deep optical imaging of the nearby radio galaxy B2~0648+27 provides further evidence for the hypothesis posed in \paperI\ that a galaxy merger formed this system. In particular the distorted optical morphology of the central region and the broad tidal arm that stretches west of the host galaxy are direct indications of a past merger event. It is not likely that a close interaction caused the observed morphology, given the lack of large, gas-rich galaxies in the vicinity of, and at approximately the same velocity as, B2~0648+27 (see \paperI).

The faint stellar tail that curls around the host galaxy appears to be settled in a partial ring, following the \HI\ ring that surrounds the system. In \paperI\ we argued that the \HI\ ring likely formed out of gas that was expelled from the merging system in large tidal tails, which (at least partially) fell back onto the newly formed host galaxy and settled into the regular rotating structure that we observe. A logical assumption is that the stellar tail/ring has a similar origin. Likely, the bulk of the stars in the ring are also tidally expelled material that, together with the \HI\ gas, fell back and settled around the host galaxy. A similar scenario has been proposed by \citet{mor97} for the elliptical galaxy NGC~5266, which also contains a giant \HI\ ring with a faint stellar counterpart. In a detailed discussion, \citet{mor97} argue that the coincidence of the faint optical stellar light with the gaseous \HI\ ring around NGC~5266 is most likely the result of a past merger event.

It is interesting, however, that peaks in the stellar distribution across the ring around B2~0648+27 appear to coincide with maxima in the \HI\ surface density. This indicates that perhaps not all the stars in the ring are tidally ejected material, but that star formation may have been triggered -- at least in localised sites or \HII\ regions -- in the tidally expelled \HI\ gas {\sl after} the merger event. In \paperI, we estimated that the highest concentration of \HI\ gas (in the eastern part of the \HI\ ring) has a surface density of 1.7 $M_{\odot}\ {\rm pc}^{-2}$, which is just below the threshold for star formation observed in galaxy disks \citep[][]{hul93,mar01}. We note, however, that the beam-size of our \HI\ observations is much larger than the resolution of our optical images. As a result, the \HI\ surface density could locally be significantly higher at the sites were star formation seems to occur. Also in other merger systems, star formation is confirmed to occur in tidal gas structures \citep[see e.g.][]{hib05}.

Obtaining reliable spectral or colour information about the stellar content of the large-scale \HI-ring (both for the diffuse stellar light and the apparent sites of star forming) is necessary to investigate in more detail whether the stars are tidal material, or formed in the tidal debris after the merger.

\subsection{Timescales of merger, starburst and AGN activity}

In \paperI\ we argued that the large mass, as well as the extended distribution, of the \HI\ gas around the host galaxy suggests that the merger was likely a major merger between two galaxy of comparable mass, of which at least one, and perhaps both, were gas-rich. This major merger must have happened roughly 1.5 Gyr ago, after which the gas and stars that were expelled in large tidal tails during the initial stages of the merger event had the time to (at least partially) fall back onto the host galaxy and settle into the regular rotating ring-like structure that we observe \citep[based on numerical simulations by][]{bar02}. The presence of a single nucleus in the morphologically distorted host galaxy provides additional evidence that the merger is in an advanced stage. The estimated time-scale for individual nuclei of the progenitor galaxies to fully merge is $1-2$ Gyr \citep[]{bar96,mih96,dim07}, which is in agreement with our estimate of the age of the merger event.

The time-delay between the initial merger and the occurrence of the major burst of star formation (as traced by the 0.3-0.4 Gyr young post-starburst stellar population in B2~0648+27) was explained in \paperI\ by the progenitor disk-galaxies already containing a prominent bulge-component \citep[see e.g.][]{mih94,mih96}. Recent numerical simulations by \citet{dim07} suggest that such a time-delay, as well as the formation of large-scale tidal tails with intense, local and clumpy star formation, may perhaps also be explained by a merger between an elliptical and a gas-rich spiral galaxy.

The spectral age of the radio source in B2~0648+27 is about 1 Myr, although it is possible that the radio source has been confined for a significantly longer time \citep{gir05a,gir07}. As discussed in detail in \paperI, there therefore appears to be a significant time-delay between the merger/starburst event and the onset of the current episode of radio-AGN activity in B2~0648+27, with the radio source appearing in the advanced stage of the merger process.

\section{Conclusions}
\label{sec:conclusions}

We presented new, deep optical B+V band imaging of the nearby radio galaxy B2~0648+27. The distorted optical morphology of the host galaxy and the presence of a broad stellar tidal arm provide further evidence for the scenario proposed in \paperI\ that a major merger event formed this system. A low surface-brightness stellar tail or partial ring follows the large-scale \HI\ ring that surrounds the host galaxy at a radius of up to 55 kpc from the centre. The gas and stars in this ring are most likely tidally expelled material that slowly fell back onto the host galaxy after the merger event, although there also appear to be sites of star formation that may have formed within the gaseous tidal debris after the merger. The observed properties of the gas and stars, as well as the time-delay between the merger and starburst event in B2~0648+27, are in agreement with simulations of a merger between two disk galaxies with a prominent bulge, or between an elliptical and a gas-rich spiral galaxy. Apparently, the triggering of the current radio source in this system was delayed with respect to the merger and starburst event.

\begin{acknowledgements}
BE thanks the the Netherlands Organisation for Scientific Research (NWO) for funding this project under Rubicon grant 680.50.0508. Data were obtained using the 2.4 m Hiltner Telescope of the Michigan-Dartmouth-MIT (MDM) Observatory, owned and operated by a consortium of the University of Michigan, Dartmouth College, Ohio State University, Columbia University and Ohio University. Many thanks to the MDM staff for their technical support.
\end{acknowledgements}

\bibliographystyle{aa} 
\bibliography{0046bib} 

\end{document}